# BUSINESS AND TECHNICAL REQUIREMENTS OF SOFTWARE-AS-A-SERVICE: IMPLICATIONS IN PORTUGUESE ENTERPRISE BUSINESS CONTEXT

Virginia Maria Araujo and José Ayude Vázquez

Informatics Department, University of Vigo, Vigo-Pontevedra, Spain

## *ABSTRACT*

*Software-as-a-Service (SaaS) is a viable option for some companies bearing their business processes. There is a considerable adoption rate, with companies already using more than two services for over two years. However, while some companies have plans to put more business processes supported by these services in the near future, others do not know if they will. They have several concerns regarding the software providers' service level. These concerns are mainly technical and functional issues, service availability and payment models. There are major changes compared to the traditional software that have implications on how the software is developed and made available to the users. The existing research addresses specific aspects and few studies give a broader view of the implications of SaaS for anyone who develops and provides software, and also for those who consumes it as an end user. What are the real needs of the Portuguese market? What fears and what is being done to mitigate them? Where should we focus our attention related to the SaaS offering in order to create more value? Thus, to analyze these questions four exploratory case studiesare used to assess the possible implications of SaaS on software developers or software providers based in Portugal and also on end-users.*

*This article appears in the context of a realistic and deep research that includes the involvement of managers, leaders and decision makers of Portuguese companies, to realize what actually constitutes a problem in SaaS and what effectively companies would like to have available in this offer. The results of this study reveal that SaaS effectively constitutes a very interesting and solid solution for the development of Portuguese companies, however there is a lack for greater efforts particularly in terms of customization for each customer (tenant) and integration with the back-end on-premise applications.*

## 1. INTRODUCTION

At the beginning of the current millennium the term SaaS is officially introduced. Although, until today, there is still no consensus about the concept of SaaS, it can be defined as a software distribution model that consists in providing a piece of software, application or service to multiple customers or tenants through the Internet (multi-tenant). The SIIA (Software & Information Industry Association) in the article entitled "Software as a Service: Strategic Backgrounder," says - "[...] In the Software as a Service model, the application or service, is deployed from a centralized data center across the network - Internet, intranet, LAN, or VPN - providing access and use on a recurring fee basis. Users "rent", "subscribe to", "are assigned" or "are granted access to" the applications from the central provider" [1].

SaaS is a very attractive alternative for small business, medium and large enterprises. The SaaS approach has the potential to transform the way the information technology (IT) departments relate to each other and even what you think about your role as a service provider [2]. There are, however, still several concerns and challenges that developers and SaaS providers have to overcome. Keep in mind that technical issues should be early identified since the're changes compared to traditional software development. Most organizations know well the value of information and how important data security is. The customers will pay for the use of the service that is configured or customized according to their specifications. This can, for example, require that their data should be separate from other customers. In this case, the provider has to ensure the full isolation that can pass through the separation of the data in its own database for that customer. Another aspect that is important to consider about the development this type of software is the business model that can be used because it can involve higher costs and larger issues on software development.

Based on this discussion and to answer the research questions we intent to establish an involvement with the managers and owners of Portuguese companies, in order to evaluate the use of SaaS in their organizations and to confirm the hypothesis that SaaS brings benefits and potentiates their development. We selected companies for the study sample to address aspects of SaaS from different points of view to have a broader and more realistic purpose of the investigation. Existing research examines specific aspects such as risk factors and adoption [3], [4] [5], and technical aspects [6] [7] [8]. A study where we take a more holistic view considering the implications in the development, provision and use of SaaS in different types of businesses (small, medium and large companies) is more than justified and even more in the Portuguese market where this research theme is still very sparing. Examine the issue from different points of view from all stakeholders, analyzing the individual aspects and interdependencies between multiple domains, allows taking a broad view of the concept and its implications for thus identify aspects and issues for future research.

## 2. LITERATURE REVIEW

Until today there is no consensus about the concept of SaaS. According to Chong & Carraro, if five people are asked about the concept of SaaS probably we get five different answers [2]. However, the experts are in agreement about some fundamental principles that distinguish SaaS from the software as a traditional product and on the other, from the simple web site. Gartner defines "cloud computing" as a style of computing where IT capabilities are massively scalable and delivered as a service to external users of technologies [9]. Cloud computing is a broader subject that cannot be confused with SaaS, which uses cloud computing only in its architecture [10]. In terms of the SaaS architecture, there are many doubts and questions about the most effective model to use. Desisto and Paquet present the four architectural models most used by SaaS developers and providers [11].

In the definition "Software deployed as a hosted service and accessed over the Internet" [2], we see that it does not describe any specific architecture for the software, it says nothing about the specific technologies or protocols to use, does not draw a distinction between service-oriented to the business and to the general consumer, and does not require specific business models. According to this definition, the main differences from SaaS are the place where the application code is executed, how it is implemented and how it is accessed. Analyzing from the general to the specific, we can identify two main categories of software as a service [12]:

a) **Line of business services,** offered to the businesses and organizations of all sizes. The lines of business solutions are usually large, customizable, aimed to make easier the processes execution such as finance, supply chain management (SCM) and customer relationship. Typically these services are provided to the customers through a service subscription.

b) **Consumer-oriented services offered to the general users**. The customer-oriented services are sometimes provided through a subscription, but are generally provided free of charge and funded through advertisements.

Desisto and Pring said that despite the expansion of SaaS, switching to this model cannot be applied in all kind of software and in all type of organization. Taking advantage of software and services, it is possible to maximize the choice, flexibility and capabilities of users in general [13]. However, before the move to a SaaS model, organizations should first obtain answers to several questions, such as [13]:

- What are the business and IT needs?
- What benefit has the IT department in adding SaaS applications to their service portfolios?
- Is there a need for customization and integration?
- What are the implications of adding hosted applications in an external environment to the organization?
- What are security levels?
- What is the roadmap of the services?

Users often face difficult choices. SaaS applications are easy to use and have lower implementation costs. SaaS providers have the advantage of lower-cost implementations compared to the traditional software. However, the thin client SaaS applications do not have the full functionality and performance of desktop systems. Moreover, it is always a challenge to integrate SaaS applications with on-premise backend systems. There are always risks of adopting a SaaS application and therefore there must always be a balance between the implementation costs and the application performance [4]. Exploring the adoption of SaaS, some studies reveal uncertainty factors, such as technical factors, process, economic risks or uncertainties associated with the functionality needs, as well as service volume. There is still a lack of research on the empirical analysis of risk factors for the delivery of SaaS applications [3]. SaaS solutions are managed and maintained by the provider, enabling customers to change the application ability and capacity without investing in new infrastructure, personnel training, or on new software licenses. The uncertainty around the service volume is no reason for the existence of incomplete contracts, since the service volume for an application can be measured using variables such as space storage, the volume of transactions and number of licenses. It is quite easy to develop metrics and set prices based on these variables, defining deployment models according to the ownership of the service [5].

According to Caldwell & Eid, although large scale businesses are less likely than small ones to consider SaaS solutions for the financial activities of GRC (Governance, Risk and Compliance), the experiences of managers show no significant drawback that prevents the use of SaaS solutions for SOX (Sarbanes-Oxley) and other types of financial activities of GRC. Companies should consider SaaS GRC solutions when [14]:

- The data that could be achieved by the provider or other third parties are not regulated information or when the regulated information is available by the provider or other third parties are contracted security services and effective access controls.

- Compliance Officers cannot readily meet the need for support and maintenance of IT infrastructure that support to GRC activities.
- The costs of a licensed application are more difficult to approve than the cost of a SaaS subscription (€10,000 per month for a SaaS subscription may be easier to justify than a € 400,000 investment to implement a licensed application plus the annual maintenance cost of 20%).

Despite the advantages of SaaS, its adoption still causes doubts in many CIOs (Chief Information Officer), especially with regard to the security and confidentiality of data. This type of question is one of the obstacles to the growth and expansion of the SaaS model, making the decision harder between either to adopt or not software as a service.

Another concern is the service availability. Donna Scott worked with Robert Desisto and Alexa Bona, to draw the attention to the unplanned downtime costs of SaaS [23].

In many cases, integrating SaaS applications with local applications, means creating dependencies that require synchronization and data transfer between the SaaS solution and one or more internal applications [15]. To answer these concerns of integration with SaaS, software providers should provide different options. Organizations have four options to solve the SaaS integration problem [16]:

- Use the APIs (Application Programming Interface) and provider's software technologies;
- Use a SaaS integration technology from a third party;
- Use integration solutions-as-a-service (IaaS);
- Use professional services or a system integrator.

SaaS customers should set their own strategies for integrating SaaS and must incorporate them in a holistic approach to multi-enterprise integration in all projects [16]. Companies should consider the flexibility and the implications of risk management in adding SaaS to their IT services portfolio. The integration and customization are critical components in the successful SaaS architecture strategies, in a service centralized IT infrastructure [15]. The main activities of a SaaS developer and SaaS provider are the development , implementation and maintenance of software that support the business processes of its customers, giving them, by this way, a greater chance of creating value. When developing applications using these models, providers should consider several factors that are not present in the common methods of providing software. The differences between the developments of software to be delivered as a good service are clear. It is necessary to consider technical aspects that change the way how this new kind of software applications are being developed [6]. The SaaS business model is the balanced configuration of various elements being essential the scalability. This is the factor that allows us to explain the difference between ASP and SaaS [7].

Offering software as a service rather than as a software product requires changes on software developers and telecom operators. It is necessary to change the thinking in three interrelated areas: business model, the software architecture and operating structure [2]. The role of the provider has to change radically from a remote application storage, to an active agent for management of a complex software ecosystem, where all the IT resources required are coordinated in order to maintain and create value for all parties involved [8]. From the literature review the conclusion is that SaaS is much more than a software distribution model and is a combination of business model, application architecture and operational structure.

## 3. METHODOLOGY

The purpose of this paper is to study the motivations and concerns of Portuguese companies regarding SaaS and identify organizational implications and techniques in the transition to this type of solutions, as well as to realize the extent to which SaaS improves their efficiency and can be a decisive factor for its growth strategy. For this, it is concluded to be more appropriate to adopt a research strategy based on a combination of research methods. We identified non-functional aspects and also technical aspects, based on the literature review and on a quantitative study results collected using a survey sent to developers, providers and SaaS end-users.

Through the combination of methods, complementing the disadvantages of one method with the advantages of the other, it increases the reliability and significance of the research. A combined approach of quantitative methods with qualitative methods complements and enriches the research [17].

The researchers feel that the biases inherent in one method may counteract the biases of other methods. From the original concept of triangulation of different data sources are being crossed quantitative methods with qualitative methods [18]. The current literature presents the following main reasons for the combination of methods:

1) The results of a method can help develop another method;
2) A method may be combined with another method to provide information on different levels or analysis units;
3) The methods may have a higher purpose and more comprehensive, changing or defending marginalized groups such as women, ethnic and racial minorities, specific communities or people with disabilities, or those who are poor;

This study uses a research strategy based on a combination of research methods sequentially. In first place a survey is used to get a broader view and to provide background information through a quantitative research, allowing subsequently to have a greater depth in qualitative research, helping to interpret and contextualize the qualitative results. Thus, the study starts with a quantitative analysis of data using two online surveys sent to a sample of Portuguese companies with the purpose of evaluating the adoption level of SaaS solutions for their organizations. One survey is sent to different types of businesses (small, medium and large) based in Portugal, developers and / or providers of SaaS solutions. A second survey is sent to companies that already adopted one or more SaaS solutions in their organizations, to collect the perspective of the users. The results of this first quantitative research complement qualitative research discussed in this paper, which aims to understand the different types of Portuguese companies according to the perspective of the decision maker, manager or director within its specific operational context and reality.

In this research, data is collected through semi-structured interviews, face-to-face with the directors and managers responsible for IT strategy of their organization or business area. The interview is a standard tool for data collection and a primary source of information for case studies [19].

Two interview guides are elaborated. One oriented to the SaaS developers and providers and another addressed to SaaS users. Each interview is audio-recorded with duration of two hours. During the qualitative interview, notes are taken from the observations made by the interviewer / researcher. Mainly, all interviews are processed in an identical manner to permit identify the existence or not differences and what factors can explain these differences.

## 4. RESULTS

The main benefits that all companies involved in this study have, in addition to the features inherent in the service that they provide, are fundamentally savings in infrastructure costs and operational maintenance that SaaS customers no longer need, and also the ease of access to the information from any location and at any time. The access to a software always updated without having to worry about installation and software deployment on their local infrastructure is another major benefit that is identified by their customers.

The change to this new business paradigm brings several implications and difficulties that companies have to overcome. It requires internal reorganizations, changes on strategies and update skills. In many cases, integrate SaaS solutions with local applications means creating dependencies that require synchronization and data transfer between the SaaS solution and one or more internal applications [15]. SaaS customers should set their own strategies for integrating SaaS and must incorporate them in a holistic approach to integrate multi-company, in all projects [16]. The integration and customization are critical components in the strategies of successful SaaS architectures and in an IT infrastructure centralized on service [15]. Integration with other applications is identified as a key requirement in many processes of any company and a point of failure in many SaaS solutions. Besides some services not making available SaaS integration processes, others always requires the software provider's intervention. Also at this point, theoretically, SaaS services provide interfaces that allow the customer to make the integration with internal systems. For this integration if at the client-side development is needed, from supply-side it needs a configuration at the most. However that has not been verified yet and even intervention in the application by the supplier is necessary. The customization is a problem found in the SaaS software. Theoretically the possibility is given to the customer to customize the application according to their specific needs without interfering with the core of the application, not impacting the processes of other customers and the correct functioning of the application. However, in practice this has not been verified. In all analyzed cases, the customization is very sparse and only gives the client the opportunity to send suggestions for improvement, entering them in the providers pipeline to be implemented in a later release available to all customers.

### 4.1. Non-functional requirements

The analysis results show that all enterprises assign a great importance to the quality of service on SaaS applications. To measure the quality level of the services associated with SaaS applications, generically the software providers are based on the security levels, availability, compliance, service levels and trust that customers place on the service that they provide.

The service reliability is closely related to the credibility of the company that provides this service has on the market. Thus, some companies establish partnerships with credible and recognized organizations in Portugal that leverage the acceptance of their SaaS offer and potentiate its development.

Cost savings for the SaaS service customer is one of the main reasons for the adoption as seen in the analyzed cases. All providers and SaaS software developers reveal that in fact the initial low investment or zero in some cases capture customers to choose this type of software. Other studies reveal that in 2014, about 34% of all new software purchases are consumed through SaaS and constitute about 14.5% of software worldwide spending [20].

Regarding the payment model, generally the model based on subscription is used. The reason for this situation of suppliers adopting subscription-based models is because it is simpler for both the provider that avoids control processes and reporting consumption of various customers, and on the other hand is also simpler for the customer who always pays the same, with no surprises in costs and highly variable. Theoretically, the subscription-based model is the most appropriate for the customer that pays exactly what it consumes. However, customer feedback, as seen in the results of the surveys and also the feedback from the interviews, is that customers prefer the subscription model that gives them more comfort.

Table 1. Non-functional requirements

| Company | Quality of service | Benefits and efficiency | Reliability | Costs saving |
|---|---|---|---|---|
| Case1 | High level | Productivity | 1) Redundancy on infrastructure, applications and data;<br>2) Without service interruption. | 1) Lower setup cost and sometimes zero;<br>2) Payable monthly, yearly, or based on use. |
| Case 2 | Provides quality | Collaboration | 1) Redundancy on infrastructure, applications and data.<br>2) Without service interruption. | 1) Communications operator does not charge any traffic in data transfer to their customers, and each customer has access to the service without additional cost.<br>2) Each accountant chamber is entitled to three subscription licenses. Above this, is paid a subscription. |
| Case 3 | Provides quality | Process improvement | 1) Credible provider with a high number of worldwide customers.<br>2) No service interruptions. | 1) Indirect costs;<br>2) Reduction of hours of classroom training;<br>3) Increase in the number of E-learning training hours. |
| Case 4 | Provides quality | Tracking the market and time-to-market | 1) Credibility of SaaS software partners;<br>2) Infrastructure support | Savings in infrastructure costs |

Table 2. Non-functional requirements (cont.)

| Company | Payment model | Time to market | Implications | Satisfaction |
|---|---|---|---|---|
| Case1 | 1) FTEs;<br>2) Subscription;<br>3) Based on use | SaaS is core competence of the company | Difficulties in accepting this type of software offer in the | Yes |
| | by transacted document;<br>4) Measurement of consumption use with online access. | | Portuguese market | |
| Case 2 | 1) Subscription;<br>2) Monthly;<br>3) Integrated in the operator's billing; | SaaS is core competence of the company | 1) Greater difficulties in software deployment without causing service interruptions;<br>2) Greater difficulties in finding qualified staff to develop a SaaS software architecture. | Yes |
| Case 3 | Subscription | 1) Cloud solution made easier the new performance management process in cross-company group;<br>2) Simpler service management. | Content for e-learning are more expensive than classroom training content. | Yes |
| Case 4 | Subscription | 1) Technological developments;<br>2) Simpler management services<br>3) Time-to-market | 1) New model of projects<br>2) Changes in the service providing<br>3) Changes in the culture of the company | Yes with improvements |

## 4.2. Technical requirements

The architecture is the foundation that determines the success in building a software system. There are many doubts and questions about what is the most effective model to use. Desisto and Paquet feature presents the four architectural models most used by developers and SaaS provider [11]:

1) Single tenant
2) Shared execution
3) Multitenant / version
4) Multitenant / Multiversion

The analyzed cases show that they have a multi-tenant application and data architecture and also multi-tenant application only. This raises some questions about the difficulties on customization in these cases.

Data security is the key in any information system not only in SaaS solutions. This is a major concern for SaaS users and one of the biggest challenges for software architects. In August 2012, Michael E. Davis presents the results of a survey conducted by InformationWeek and aimed to assess the extent to which providers of cloud services ensure the security of data customer and what the risks associated with cloud services [21].

Much remains to be done on these topics that are requirements to consider in the design of any software application and also on SaaS applications. These are the challenges for engineers, software architects and researchers, whose mission is to transform these challenges into opportunities and solutions to these problems. Some initiatives and projects are ongoing to investigate these issues for companies to create new products, new value that can address these concerns. The European Union under the Europe Digital agenda program is conducting a great political movement to research and find solutions to these problems. The new strategy of the European Commission entitled " Unleashing the potential of cloud computing in Europe " outlines actions to achieve a net gain of 2.5 million of new jobs in Europe, and an annual increase of EUR 160 billion to EU GDP in 2020 [22] . Under the Europe 2020 strategy, a project titled " Tcloud " is created financed by funds of the European Union and the FP7 (Seventh Framework Programme), for which Portugal contributes with the participation of some companies.

Security has been one of the most criticized aspects of the solutions in the Cloud and on which projects have been developed further investigation. The analyzed cases show that this is a constraint to SaaS adoption compared to on-premise solutions. They use authentication processes, SSL certificates and encryption.

Availability is another of the aspects we consider of utmost importance in the SaaS solutions that must be available whenever a customer needs it. So, have a disaster recovery processes and replication, which in addition to allowing recover from disasters as they have replication between data centers that are in different geographical locations, allows them to make interventions on the platform without causing unavailability of the service.

The virtualization technology used by some companies greatly facilitates this process. As development technologies they use Java, .Net, Web services, SOA, Mobile, Open Source, Postgres, C, Ruby On Rails, HTML 5.

Table 3. Technical requirements

| **Company** | **Functionality levels, updates and roadmap** | **Customization** | **Integration with other applications** | **Architecture** | **Security** |
|---|---|---|---|---|---|
| **Case 1** | Releases Monthly and quarterly | Few | Yes | Multi-tenant aplication and data | 1) Encryption<br>2) Authentication |
| **Case 2** | Frequent updates | No | Partly | 1) multi-tenant applicati | 1) Encryption<br>2) Authentic |
| | | | | on<br>2) single-tenant database | ation |
| **Case 3** | 1) Three Releases per year<br>2) Annual releases | No | Yes | | 1) Encryption<br>2) Authentication |
| **Case 4** | Annual releases | Few | Few | Multi-tenant aplication and data | 1) Encryption<br>2) Authentication |

Table 4. Technical requirements (Cont.)

| **Company** | **Availability** | **Virtualization** | **Technologies** | **Emerging trend in the software industry** |
|---|---|---|---|---|
| **Case 1** | 1) Disaster Recovery<br>2) Without interruption | Yes, Vmware | Java,<br>Net,<br>Webservices,<br>SOA,<br>Mobile | Globalization |
| **Case 2** | 1) Disaster Recovery<br>2) Without interruption | Yes, Vmware | Open Source,<br>Linux server,<br>Postgres Database, C,<br>Java<br>Ruby On Rails<br>HTML 5 | Mobility |
| **Case 3** | 1) Disaster Recovery<br>2) Without interruption | Doesn't know | Cornerstone | 1) More services in the cloud with modularity and more integrated;<br>2) Greater customization can direct the investment |
| **Case 4** | 1) Disaster Recovery<br>2) Without interruption | Yes | Dynamics NAV,<br>Dynamics CRM,<br>Business Objects<br>Sana Commerce | 1) Mobility<br>2) SaaS architectures standardization |

## 5. DISCUSSION AND IMPLICATIONS FOR PORTUGUESE COMPANIES

All selected managers consider that the Software-as-a-Service brings benefits and makes the management of its business easier, although varying systems and technology used as well as business processes and needs. The savings in infrastructure and operational costs, the access to software always updated, always available from anywhere and any device, the ease of software deployment and centralized management, are the major advantages and major benefits cited by all interviewed companies. The service reliability and information data security are referred as the major concerns for all inquired SaaS companies. The customization and integration with other systems that support the business processes of different companies, are the major problems in the adoption of SaaS and therefore do not consider its use in critical business processes.

In contrast to previous studies, this study reveals the adoption of SaaS for medium and large companies on the predictions of micro and small businesses that would be interested in this model of providing Software. The study find outs that generally, the developers do not have a software architectural model sufficiently mature to allow the coexistence of different organizations with specific functionality within the same application in a centralized application environment. In addition, to the users concerns regarding information data security, as evidenced by several previous studies, it is confirmed that this is the main constraint to SaaS adoption leading to greater research and more technological developments.

The results imply that managers of Portuguese companies recognize the added value of the Software-as-a-Service in their organizations. However, there are some modifications to do, to adapt and evolve. Listed below are the findings of this study and its implications:

- It is confirmed that in fact the SaaS model can turn into a positive way to develop and deploy the software. It is necessary for developers and software providers reconsider SaaS in their offer;
- SaaS is directly related to volume, scalability and speed. The mass adoption is based on the experience and satisfaction. SaaS developers and providers must have agile ways of gathering customer feedback and measure their satisfaction level.
- The scalability of the solutions is crucial to achieve critical volumes, in order to generate the desired profits. It is necessary to predict the needs for scalability and draws the SaaS solution accordingly to those needs. It is necessary to deep the design of highly scalable solutions that can easily increase or decrease according to the needs of SaaS software use.
- It is important to design SaaS applications in order to make it easy to adapt with the needs of each customer or organization. The recognition by the user that the software works for their organization and their specific needs starting the virtuous cycle of references that is reflected in a viral marketing. This is important for customer retention and for the continuity of the profits. The customization is a confirmed need for all organizations and as such it is necessary to explore the software architectures.
- The security and privacy of information data is an area of concern and uncertainty about the use of SaaS. The focus on application architecture, ensuring security information has to be a strong point of attention by the software developers. Customer data and payment information in e-commerce solutions is imperative to be transacted securely. The SaaS offer should highlight this point and show to potential customers.
- The inquired companies are focused on Time-to-Value. There should be a short distance between the decision and the implementation of an application or feature. SaaS is a rapid application deployment.
- SaaS is advancing rapidly. The time-to-market is more important than on-premise solutions. The integration of SaaS applications with on-premise solutions is a necessity.
- The SLA (Service Level Agreement) of SaaS services is very important. Users are concerned with the service levels over which they have control. So, it is very important to have a transparent view of the service and its performance. SaaS providers should therefore supply reports of service performance.

- The complex licensing models can be a handicap (disadvantage). Because SaaS is different from traditional software sales plan, it is imperative to develop sales plans and set appropriate licensing models that must be simple to understand.
- Companies have a valid concern between having too much infrastructure and have little infrastructure. Some of them think that a large infrastructure slows down the organization and inhibits innovation. Others think that no infrastructure is unlikely to succeed and climb. Maintaining an infrastructure in-house using SOA principles and Private Cloud used as support for SaaS solutions, can be considered viable for most organizations.

## 6. CONCLUSION AND CONTRIBUTIONS

The results of this study imply that managers of Portuguese companies recognize the added value of the Software-as-a-Service in their organizations. However, some changes are needed, to adapt and evolve. SaaS is not just a different software distribution model but a new way of doing business with software. Today, companies have at their fingertips the technology that allows them to embrace new markets and new business. They should identify business strategies, needs of organizational changes and human resources such as acquiring new skills and reorganization of functions and tasks.

The major contributions of this study are the knowledge of the implications that SaaS can have in the competitiveness enhancing and development of Portuguese companies to carefully plan appropriate investments in its human capital, processes and relevant technologies. The non-functional requirements presented in a business perspective and technical requirements presented in a technological perspective are based on an extensive study. The added value of this study lies mainly to the integration of these two perspectives in terms of who provides and who uses the software. Requirements to evaluate the impact of a SaaS transition and provide an understanding of the Portuguese company's real needs that can be generalized to other geographies. The reported research results and the raised issues can provide information for decision making regarding the adoption of SaaS in the strategic lines of IT organizations. It can also be useful in the discussion or clarification of technological issues needed to support the information data and business processes.

This study brought new insights to conduct future studies to answer the following questions and recommendations:

- The way companies use SaaS services need to be better understood. Does it really take advantage of them? Or the only reason for adoption is the lower setup cost?
- How to solve the integrating problems of SaaS? How to integrate SaaS services with on-premise applications? How to create an efficient platform for integrating SaaS?
- What factors influence the implementation of SaaS applications in an organization? What factors determine the type of solution on-demand, on-premise, or mixed?
- What SaaS architecture proposals can be used to create highly efficient solutions of Meta-Data, Multi-Tenancy, easily adaptable to each organization and country?
- How to extend the identification of differences between software on-demand and on-premise, in order to define a methodology for developing SaaS applications, considering development techniques and agile methodologies such as Scrum, eXtreme Programming and Lean software development (LSD)?
- This study shows some reluctance and uncertainty about the adoption of SaaS by the Portuguese companies. We suggest new studies and the development of new tools to assess more realistically the causes of this insecurity.

- Opportunities for further research arise from methodological limitations, expanding the scope of the study based on the results obtained. A future study might adopt a more rigorous positivist approach to collect data from a larger number of companies using propositions or hypotheses generated from the results of this study.

## REFERENCES

[1] SIIA (2001). Software as a service: strategic backgrounder s.l.: Software and Information Industry Association, p.4.

[2] Chong, Frederick, Carraro, Gianpaolo (2006). Architecture Strategies for Catching the Long Tail. Microsoft Corporation, p.1. Accessed from http://msdn.microsoft.com/en-us/library/aa479069(printer).aspx

[3] Benlian, A., and Hess, T. 2010. "The Risks of Sourcing Software as a Service - an Empirical Analysis of Adoptors and Non-Adopters," 18th Eurpean Conference on Information Systems, T. Alexander, M. Turpin and J. van Deventer (eds.), Pretoria, South Africa.

[4] Fan, M., Kumar, S., and Whinston, A.B. 2009. "Short-Term and Long-Term Competition between Providers of Shrink-Wrap Software and Software as a Service," European Journal of Operational Research (196:2), pp. 661 - 671.

[5] Xin, M., and Levina, N. 2008. "Software-as-a Service Model: Elaborating Client-Side Adoption Factors," 29th International Conference on Information Systems, R. Boland, M. Limayem and B. Pentland (eds.), Paris, France.

[6] Espadas, J., Concha, D., and Arturo, M. 2008. "Application Development over Software-as-a-Service Platforms," The Third International Conference on Software Engineering Advances, H. Mannaert, T. Ohta, C. Dini and R. Pellerin (eds.), Sliema, Malta, pp. 97-104.

[7] Luoma, E., and Rönkkö, M. 2012. "Software-as-a-Service Business Models," Communications of Cloud Software (1:1).

[8] Sääksjärvi, M., Lassila, A. and Nordström, H. 2005. "Evaluating the Software as a Service Business Model: From Cpu Time-Sharing to Online Innovation Sharing," IADIS International Conference e-Society 2005, P.K.a.M.M.-P. Petro Isaisas (ed.), Qawra, Malta, pp. 177-186.

[9] Gartner (2013). Gartner IT Glossary - Cloud Computing. Accessed from http://www.gartner.com/it-glossary/cloud-computing/

[10] Desisto, Robert, Plummer, Darly, Smith, David (2008). Tutorial for Understanding the Relationship Between Cloud Computing and SaaS, S.l. Gartner, ID: G00156152.

[11] Desisto, Robert, Paquet, Raymond, (2006). Consider Three Differentiating Attributes of SaaS. S.l. Gartner, ID: G00141606.

[12] Choudhary, Vidyanand (2007). Software as a Service: Implications for Investment in Software Development. IEEE IT Professional, 40th Hawaii International Conference on System Sciences, HICSS 2007, p.209.

[13] Desisto, Robert, Pring, Ben (2008). Essential SaaS Overview and Guide to SaaS Research. S.l. Gartner, ID: G00158249.

[14] Caldwell, French, Eid, Tom (2007). Is SaaS Safe for Financial Governance, Risk and Compliance Solutions? S.l. Gartner, ID: G00150913

[15] Chong, Frederick, Carraro, Gianpaolo, Wolter. R. (2006). Multi-Tenant Data Architecture. MSDN Library, Microsoft Corporation. http://msdn.microsoft.com/en-us/library/aa479086(printer).aspx

[16] Lheureux, Benoit (2008). SaaS Integration: How to Choose the Best Approach. S.l. Gartner, ID Number: G00161672.

[17] Hayati, D, Karami, E., Slee, B (2006). Combining qualitative and quantitative methods in the measurement of rural poverty. Social Indicators Research, Springer, pp.361-394.

[18] Creswell, J.W. (2009a). Research design: Qualitative, Quantitative, and Mixed Methods Approaches. Thousand Oaks, Calif.: SAGE Publications, pp.18.

[19] Yin, Robert. K. (2003). Case Study Research: Design And Methods. Sage Publications, Inc., 5. ISBN 978-1-4129-6099, pp. 11.

[20] Mahowald, Robert (2010). Worldwide Software as a Service 2010-2014 Forecast: Software Will Never Be the Same. s.l. IDC, ID Number: 223628

[21] Davis, Michael A. (2012). Cloud Security and Risk Survey. InformationWeek, Aug 2012. Report ID: R5080812.
[22] Europe (2013). Digital Agenda: New strategy to drive European business and government productivity via cloud computing. Accessed from http://europa.eu/rapid/press-release_IP-12-1025_en.htm
[23] Scott, Donna, Bona, Alexa, Desisto, Robert (2006). Beware of Unplanned Downtime When Using Softwareas-a-Service Providers. S.l. Gartner, ID: G00140952

## Authors

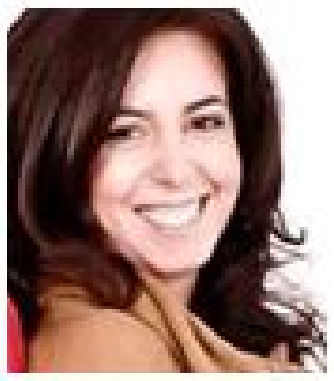

**Virginia Maria Araujo**: Received her Licentiate degree in Computer Sciences in 1995 from Minho University, Portugal, her MAS degree in 2009 from Vigo University, Spain, in Computer Engineering. She works as a Senior Project Manager in a big Portuguese company and at the same time she is a Ph.D.student in Vigo University, Spain. Her interest subjects include Cloud computing, Software Engineering, IT Technologies, Data Security, and Project Management.

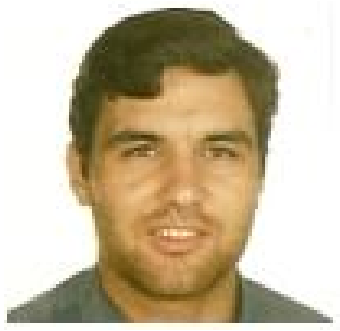

**José Ayude Vázquez**: Received his Ph.D. degree in computer science from Vigo University, Spain, in 2001, his MSc from Vigo university, Spain, in 2011. He is currently an associate professor in the Department of Informatics of Vigo University, Spain. His research interests include Software Engineering, Object Oriented Programming and Content Analysis Software.